# p$^+$-Al$_{0.3}$Ga$_{0.7}$Sb Pocket-Implanted L-shaped GaSb/InAs Heterojunction Vertical n-TFETs

Bhupesh Bishnoi[1] and Bahniman Ghosh[1, 2]

[1]DEPARTMENT OF ELECTRICAL ENGINEERING, INDIAN INSTITUTE OF TECHNOLOGY, KANPUR, 208016, INDIA

Email: bbishnoi@iitk.ac.in

[2]MICROELECTRONICS RESEARCH CENTER, 10100, BURNET ROAD, BLDG. 160, UNIVERSITY OF TEXAS AT AUSTIN, AUSTIN, TX, 78758, USA

Email: bghosh@utexas.edu

ABSTRACT

In the present work, we have investigated the performances of p$^+$-Al$_{0.3}$Ga$_{0.7}$Sb Pocket-Implanted L-shaped GaSb/InAs staggered-bandgap (SG) heterojunction vertical n-channel tunnel field-effect transistors (TFETs) of 4 nm thin channel structures with the gate lengths of 20 nm. In this L-shaped nonlinear geometry the gate electric field and tunnel junction internal field are oriented in same direction. We have used a 3-D full-band atomistic $sp^3d^5s^*$ spin-orbital coupled tight-binding method based quantum mechanical simulator. We have investigated current–voltage characteristics, ON-current, OFF-current and subthreshold swing as functions of equivalent oxide thickness ($T_{ox}$), gate length ($L_G$), drain length ($L_D$), gate undercut ($L_{UC}$), dielectric constant and drain thickness ($T_{InAs}$) in nonlinear L-shaped TFETs for low subthreshold swing and low voltage operation.

# INTRODUCTION

In present scenario Tunnel field-effect transistors (TFETs) are prospective candidates due to their steep subthreshold swing (SS), better $I_{ON}$ to $I_{OFF}$ current ratio and high drive current at low voltage operation. Hence, TFETs will reduce the overall power consumption of nanoelectronics integrated circuits.[1] TFETs are able to break the fundamental limit of subthreshold swing of 60 mV/decade because in TFETs charge carriers are injected into the channel by quantum-mechanical band-to-band tunneling (BTBT) of valence band electrons instead of thermionic emission process of conventional MOSFETs. Hence, SS can go below 60 mV/ decade. Historically, the evolution of tunnelling device was first proposed by Quinn *et al.* in 1978 which is the gated p-i-n structure.[2] However, TFETs have low $I_{ON}$ current, which reduces speed response of TFETs based circuit. But, carefully designed electrostatically optimized device structure can improve $I_{ON}$ current. As per the 2012 ITRS roadmap the future target parameters for TFETs are: $V_{DD}$ is less than 0.5 V, 100 milli-amperes of $I_{ON}$, $I_{ON}/I_{OFF} > 10^5$ and SS far below 60mV per decade. A steep slope and high tunnelling current can be achieved if minute change in gate voltage can change source tunnelling barrier's transmission probability from zero to unity.[2] To achieve this, instead of lateral TFETs geometries, vertical transistor geometries are proposed to enable lower off-state leakage, high drive current with better gate electrostatics control. In vertical transistor geometries the tunnel junction is oriented such that tunnel junction internal field and gate field are aligned with each other. [3] To increase the $I_{ON}$ and hence the tunnelling probability one seeks for efficient ways to reduce the bandgap. Materials with staggered-bandgap (SG) alignment e.g. $Al_{0.3}Ga_{0.7}Sb$/InAs are promising to build TFETs because of a small band overlap and narrow band gap of 0.73 eV and 0.36 eV for GaSb and InAs respectively. The combination of $Al_{0.3}Ga_{0.7}Sb$ and InAs material gives rise to type-II staggered-bandgap alignment (i.e., conduction band minimum of one material lies in between the conduction band minimum and valence band maximum of other material) which inherently favors tunnelling and leads to low resistivity in the junction. Moreover lighter effective masses also enhance the probability of tunnelling for InAs and GaSb. [4] In the heterostructure TFETs, the materials are chosen such that on one side of junction the material has a smaller bandgap and hence width of the energy barrier at that side of junction can be decreased in the ON-State while on the other side of junction, the material has a larger bandgap which creates the energy barrier width of largest possible size at that side of junction and keeps the off state current low. Group III–V materials have another advantage of lattice-matched growth during growth processing. [2] In this article, we demonstrate various electrostatic and geometrical considerations like equivalent oxide thickness ($T_{ox}$), gate length($L_G$), drain length($L_D$), gate undercut ($L_{UC}$), dielectric constant and drain thickness ($T_{InAs}$) that influence the scaling and design of $p^+$-$Al_{0.3}Ga_{0.7}Sb$ Pocket-Implanted L-shaped GaSb/InAs staggered-bandgap (SG) heterojunction vertical n-TFETs. Recently vertical heterojunction TFETs are experimentally demonstrated in Ref. [4, 5]. In the vertical L-shaped geometry tunnelling barrier width is optimized by applying the gate modulated electric field as the electric field and tunnelling path is aligned in the structure. Hence, in the design we have to overlap the gate on the tunnelling region as source region is covered by the top gate. By this design practice a factor of 10 improvement is visible in ON-current and steep subthreshold swing can be achieved.[6] In commercial two dimensional device simulators like Synopsys TCAD, tunnelling models are used based on Wentzel-Kramers-Brillouin (WKB) approximation. These models are 1-D in nature and neglect confinement effects and band quantization effects. Since in our device structure, the electrostatics and current path are 2-D, for more accurate analysis, we must go beyond the 1-D tunnelling models. In the source region quantization effects influenced the effective barrier height for tunneling and as a result density of states and absolute current level in the channel is changed. [7, 8] For that purpose we have used a 3-D full-band atomistic *$sp^3d^5s^*$* spin-orbital coupled tight-binding method based quantum mechanical simulator. [9-12] We have investigated the performances of $p^+$-$Al_{0.3}Ga_{0.7}Sb$ Pocket-Implanted L-shaped GaSb/InAs staggered-bandgap (SG) heterojunction vertical n-TFETs of 4 nm thin channel structures with the gate lengths of 20 nm. A full self-consistent quantum mechanical simulation including electron-phonon scattering can in principle describe TFETs with high accuracy. [13-16] However, it requires an extremely high computational intensity to solve such NEGF equations. [17] Usually coherent transport simulations are performed to obtain the upper device performance limit. Various local and nonlocal BTBT models are incorporated in the commercial device simulator but they do not address quantization effects. [18] Intensive research work is going on in TFETs as power-supply scaling below 0.5 V is possible in these devices and at low voltages TFETs can outperform aggressively scaled MOSFETs. [19, 20] Steep subthreshold swing can be achieved by in line field orientation of Gate field and tunnel junction internal field as analytically formulated by Zhang *et al.* [21] On the basis of same concept Hu *et al.* proposed a new geometric configuration of pocket TFETs and Salahuddin and Ganapathi also proposed similar TFETs [22, 23] and Ford *et al.* experimentally realized these devices [24]. Asra *et al.* reported the Vertical Si homojunction TFETs [25] and Agarwal *et al.* reported the InAs homojunction TFETs. [26] Very recently Quantum Mechanical Performance of Pocketed Line Tunnel Field-Effect Transistors is reported by Verreck *et al.* as for hetero junction device no analytical model and theory exists for BTBT tunnelling in between an indirect and a direct semiconductor. [27]

## II. DEVICE STRUCTURE

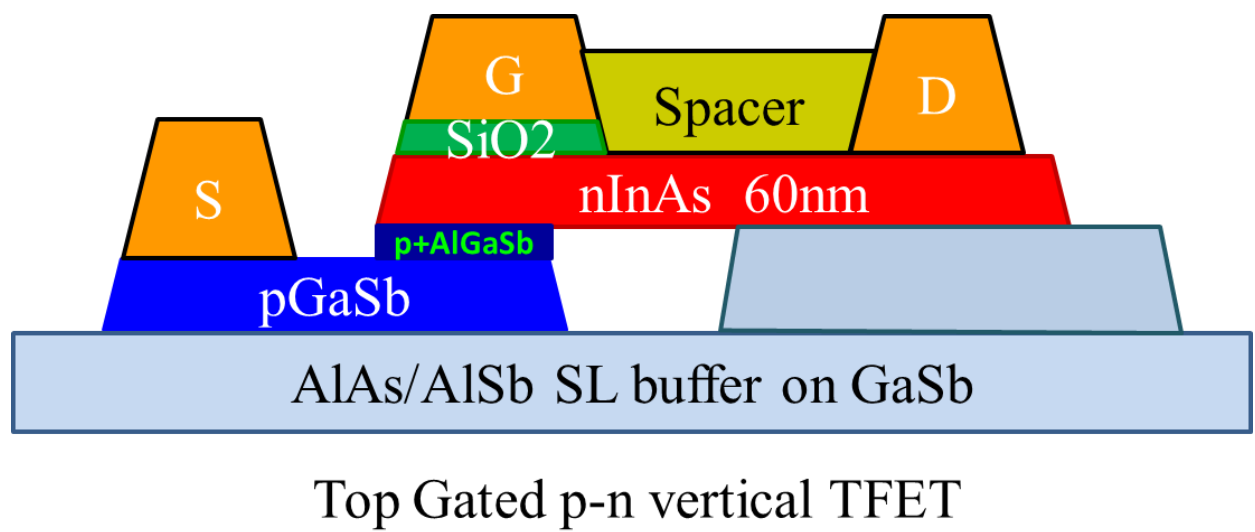

Fig.1. Cross section cartoon diagram of the simulated p$^+$-Al$_{0.3}$Ga$_{0.7}$Sb Pocket-Implanted L-shaped GaSb/InAs staggered-bandgap (SG) heterojunction vertical n-TFETs device

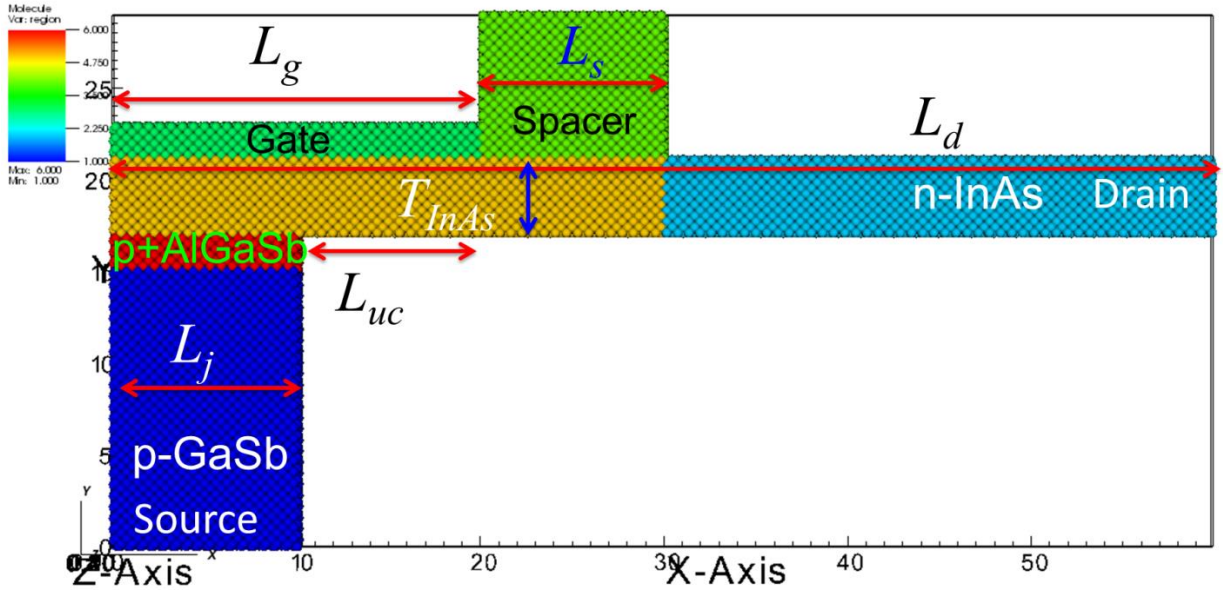

Fig.2. Atomic Structure and Geometry of $p^+$-$Al_{0.3}Ga_{0.7}Sb$ Pocket-Implanted L-shaped GaSb/InAs staggered-bandgap (SG) heterojunction vertical n-TFETs

In the figure 1 we have shown cross section cartoon diagram of simulated $p^+$-$Al_{0.3}Ga_{0.7}Sb$ Pocket-Implanted GaSb/InAs TFETs device structures. The heterostructure $p^+$-$Al_{0.3}Ga_{0.7}Sb$ Pocket-Implanted GaSb/InAs TFETs consists of 4 nm n-type InAs channel with a doping density of $5\times10^{17}$ cm$^{-3}$ and a 10 nm p-type GaSb source with doping density of $4\times10^{18}$ cm$^{-3}$. In between source and drain region we have $p^+$-type $Al_{0.3}Ga_{0.7}Sb$ Pocket-implanted source injector with doping density of $4\times10^{19}$. N-type InAs has a band gap of 0.354 eV and p-type GaSb has a band gap of 0.7266 eV. The source and drain regions measure 10nm and 60nm in length, respectively. The drain thickness $T_{InAs}$ is set to 4 nm. A $SiO_2$ gate thickness of 1.9 nm is used initially and then varies to 4 nm. The gate metal length defined the gate length ($L_G$) of device. To fulfill the ITRS requirement we started our simulation with 20 nm gate length ($L_G$). We also used $HfO_2$ high-K dielectric as gate material for another study. Between the drain and gate contact $SiO_2$ spacer of an overlap length ($L_S$) is used. Overlap spacer decoupled the gate–drain region and reduced the ambipolar conduction. The drain length ($L_D$) is defined by the length of the drain and junction length ($L_J$) is length of active tunnelling junction and hence gate overlaps the junction length. Additionally the InAs channel has an undercut of a length ($L_{UC}$) which is required to achieve a steep SS. Type-II (staggered-bandgap) band alignment of L-shaped $p^+$-$Al_{0.3}Ga_{0.7}Sb$ Pocket-

Implanted GaSb/InAs TFETs is shown in the figure 3. Direction of transport in the channel is along <100> crystal axis and orientation of surface is along (100). Due to quantization effect in junction length ($L_J$) smaller than 10 nm SG characteristic may vanish at SG heterostructure. However, we can shift the overall band structure by the tight-binding onsite energy to obtain different band alignments [28] GaSb/InAs heterojunction staggered-bandgap structure has another advantage that transport is by mixture of electrons/holes.

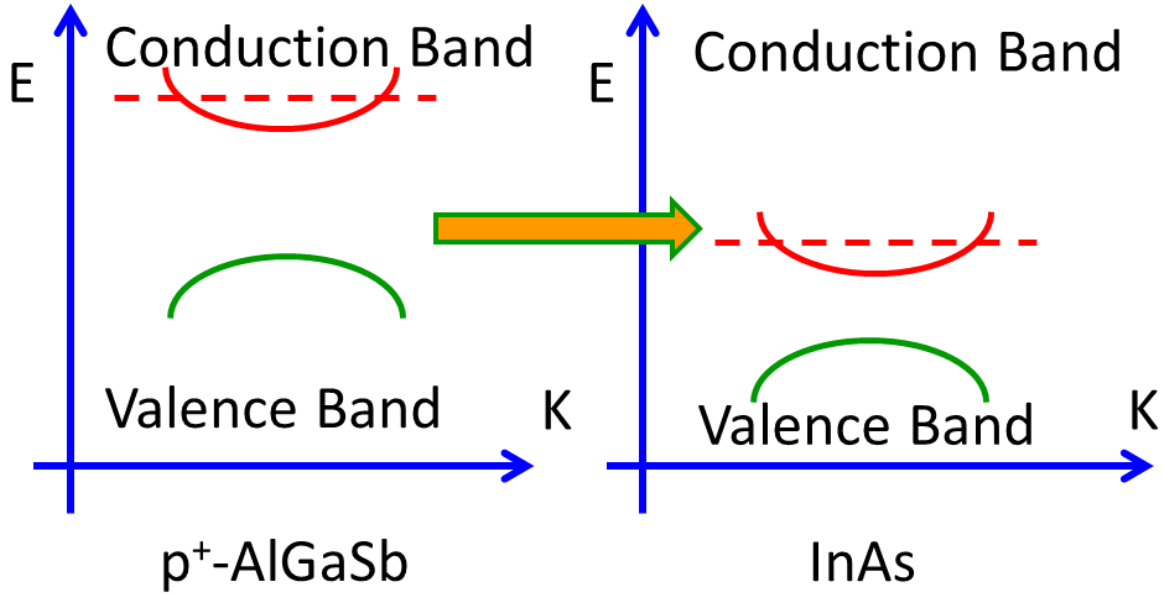

Fig.3. type-II (staggered-bandgap) band alignment of L-shaped $p^+$-$Al_{0.3}Ga_{0.7}Sb$ Pocket-Implanted GaSb/InAs TFETs

SIMULATION APPROACH

To study 2D electronic transport in L-shaped nonlinear geometry, we have used a 3-D full-band atomistic $sp^3d^5s^*$ spin-orbital coupled tight-binding method based quantum mechanical simulator which solves Schrödinger and Poisson equations self-consistently for potentials and Local Density of state (LDOS) in Non-Equilibrium Green Function (NEGF) formalism.[29-32] The advantage of this method is that it can handle arbitrary geometries and complicated 2D structures like Band-to-Band tunnelling (BTBT) device. NEGF quantum transport method gives output in terms of potential and space charge in 3D geometry; energy resolved transmission and Spectral function, Local Density of state of electrons/ holes in space and energy. In the present work, a method combining the semiclassical density and Non-equilibrium Green's Function (NEGF) formalism is used to achieve an efficient simulation of $p^+$-$Al_{0.3}Ga_{0.7}Sb$ Pocket-Implanted GaSb/InAs TFETs. [33-36] This method assumes an equilibrium carrier distribution in device and states are filled according to quasi-Fermi levels in the triangular quantum well, which could vary spatially and mimic strong scattering. Taking one unit cell from lead and calculating the bandstructure in the transport direction, we obtain the effective bandgap after confinement. The effective mass is calculated from the doping density and doping degeneracy in the contacts. The doping degeneracy is calculated using the $sp^3d^5s^*$ tight-binding model self-consistently for the same unit cell of TFETs with equilibrium boundary condition. The modification of density of states due to confinement is included in the process. We used the quantum transport simulator which is multi-dimensional, massively parallel, based on atomistic $sp^3d^5s^*$ spin-orbital coupled tight-binding representation of the band structure. Quantum transport simulation can be done either in the ballistic transport regime by Wave Function formalism or in diffusive transport regime with scattering mechanism by Non-equilibrium Green's Function formalism (NEGF). Wave Function (WF) formalism is computationally efficient but numerically equivalent to the Non-equilibrium Green's Function formalism (NEGF). [37-38] In WF formalism, we are solving sparse linear systems of equations while in comparison in NEGF formalism we are solving matrix inversion problems. In the transport calculation 960 energy points and 31 momentum points along the transport path are considered. In the active region of device every atom is represented by a matrix. In our simulation domain 12726 is total number of atoms and 11646 are number of active atoms taken into account in the Schrödinger equation. There are 542 surface atoms which are used in Poisson equation. The overall Hamiltonian matrix of the system is the tri-diagonal block matrix of sparse blocks. Number of atoms in each atomic layer will decide size of the sparse blocks. Gate dielectric layer and spacer layer are modeled as imaginary materials layer which has infinite bandgap as they separate the gate contacts and InAs channel and do not participate in transport calculation. Hence in the Poisson equation they are characterized by their relative dielectric constant. In the figure 4 flow of Quantum Transport simulation for L-shaped $p^+$-$Al_{0.3}Ga_{0.7}Sb$ Pocket-Implanted GaSb/InAs TFETs is shown. Into the device structures holes and electrons are injected by the drain and source contacts with different wave vectors and energies and carriers charge densities and current densities are obtained.

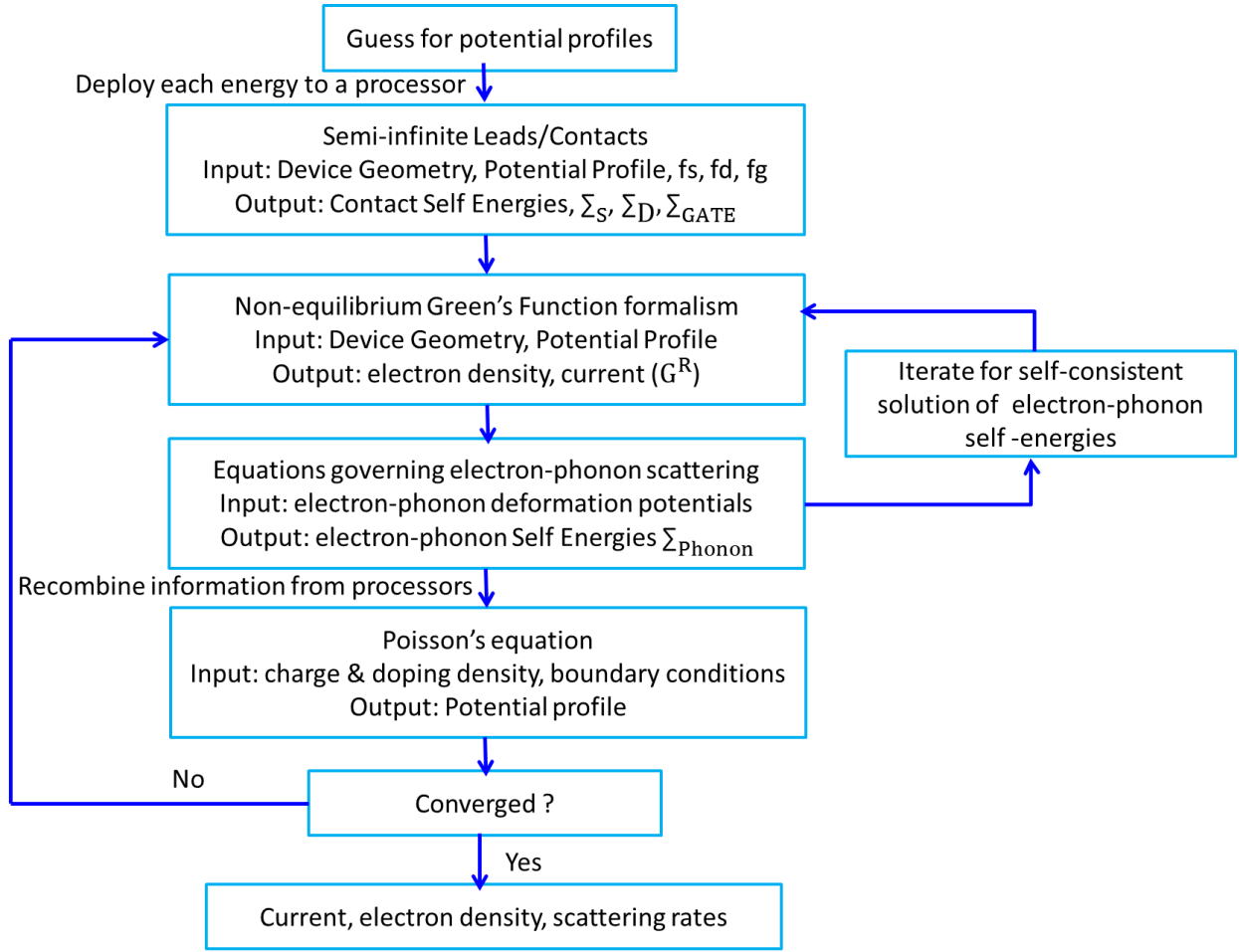

Fig.4. Flow of Quantum Transport simulation for L-shaped $p^+$-$Al_{0.3}Ga_{0.7}Sb$ Pocket-Implanted GaSb/InAs TFETs

Electrons tunnel from valence band of $p^+$-$Al_{0.3}Ga_{0.7}Sb$ to conduction band of n-InAs and these tunneling electrons contribute to both charge densities and current. These carrier charge densities are coupled self-consistently to the calculation of electrostatic potential. Hence, the solutions of the Poisson and Schrödinger equations are parallelized. [39] In the simulator the band gap, effective masses and conduction and valence bands of various semiconductor materials are modeled with spin-orbit coupling and it also incorporates quantization effects due to narrow size effect. Neglecting these effects will increase the band gap and underestimate band-to-band tunnelling probability. Atomistic $sp^3d^5s^*$ spin-orbital coupled tight-binding representation of energy band gap also accounts for the imaginary band dispersion. Hence, even in the absence of scattering process BTBT processes are accurately modeled for direct bandgap materials. It also incorporates phonon-assisted band-to-band tunnelling. [40]

RESULTS

A. energy-position resolved local density of states LDOS (x, E) and energy-position resolved electron density spectrum $G_n$ (x, E) distribution

Figure 5 shows the energy-position resolved local density of states LDOS (x, E) and energy-position resolved electron density spectrum $G_n$ (x, E) distribution of the $p^+$-$Al_{0.3}Ga_{0.7}Sb$ Pocket-Implanted GaSb/InAs staggered-bandgap vertical TFETs at $V_{DS}$= 0.3V in the ON-state conditions with variation of $V_{GS}$ from -0.1 V to 1.2 V with the step voltage of 0.1 V. In this ON-state biasing condition, gate modulates the position of the channel barrier which is due to the (staggered-bandgap) energy band and hence channel conduction band is pulled down below the source valence band to increase the source injection. The ON-state is clearly visible in the Figure 5. It also shows the heterojunction staggered-bandgap at the source–channel interface. In the Figure 5 we also showed the energy-position resolved electron density spectrum $G_n$ (x, E) distribution which shows the occupation of LDOS (x, E) by the respective source and drain contact Fermi reservoirs. Energy-position resolved electron density spectrum is shown on log scale in the $p^+$-$Al_{0.3}Ga_{0.7}Sb$ Pocket-Implanted GaSb/InAs TFETs at $V_{DS}$= 0.3 V. Figure 5 also shows original (staggered-bandgap) band alignment and amount of band shift in SG characteristic due to quantization.

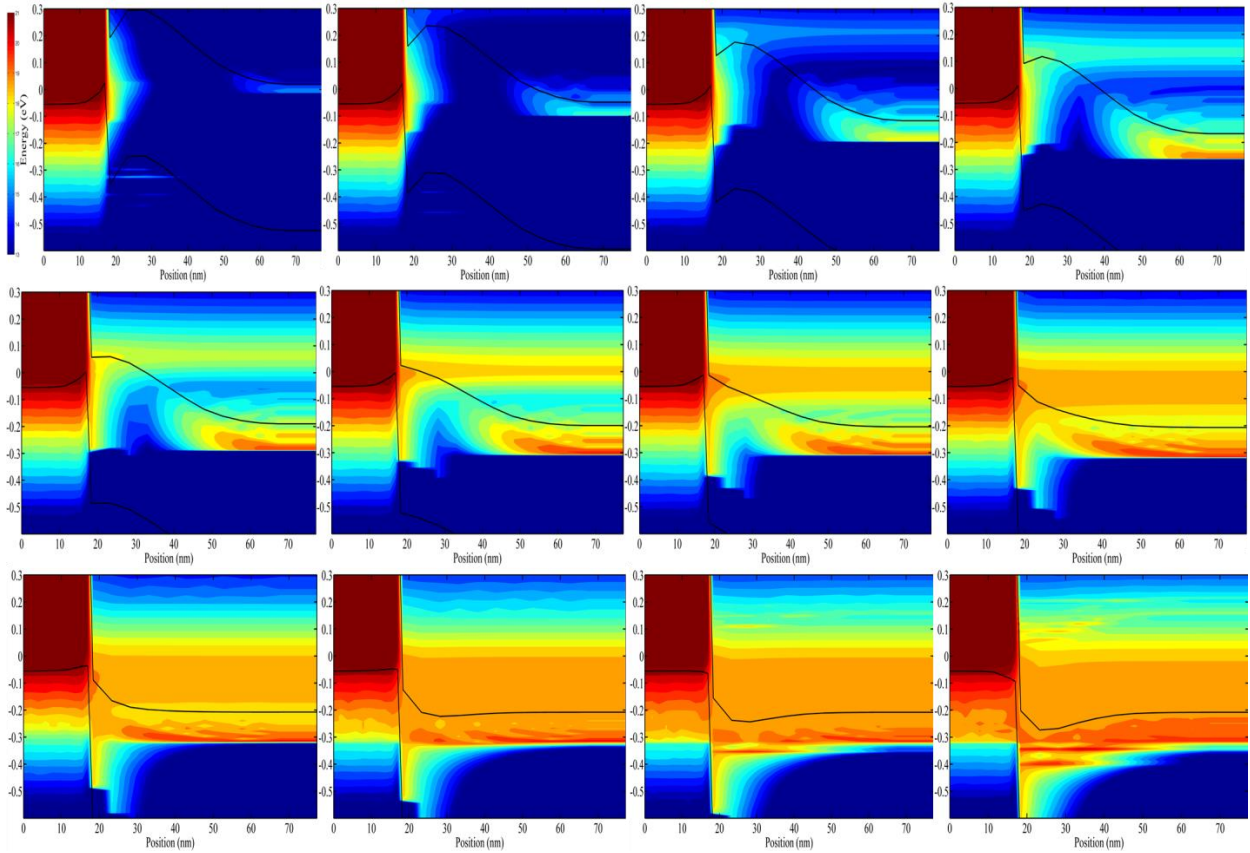

Fig.5. Energy-position resolved LDOS (x, E) and energy-position resolved electron density spectrum $G_n$ (x, E) distribution in the ON-state $p^+$-$Al_{0.3}Ga_{0.7}Sb$ Pocket-Implanted GaSb/InAs TFETs

Figure 6 shows the energy-position resolved local density of states LDOS (x, E) and energy-position resolved electron density spectrum $G_n$ (x, E) distribution of the $p^+$-$Al_{0.3}Ga_{0.7}Sb$ Pocket-Implanted GaSb/InAs TFETs at $V_{DS}$= 0.03 V in the OFF-state conditions with variation of $V_{GS}$ from -0.1 V to 1.2 V with the step voltage of 0.1 V. OFF-state leakage current is mainly due to Phonon absorption assisted tunnelling current. In the drain carrier thermalization due to Phonon emissions is also shown in the figure 6.

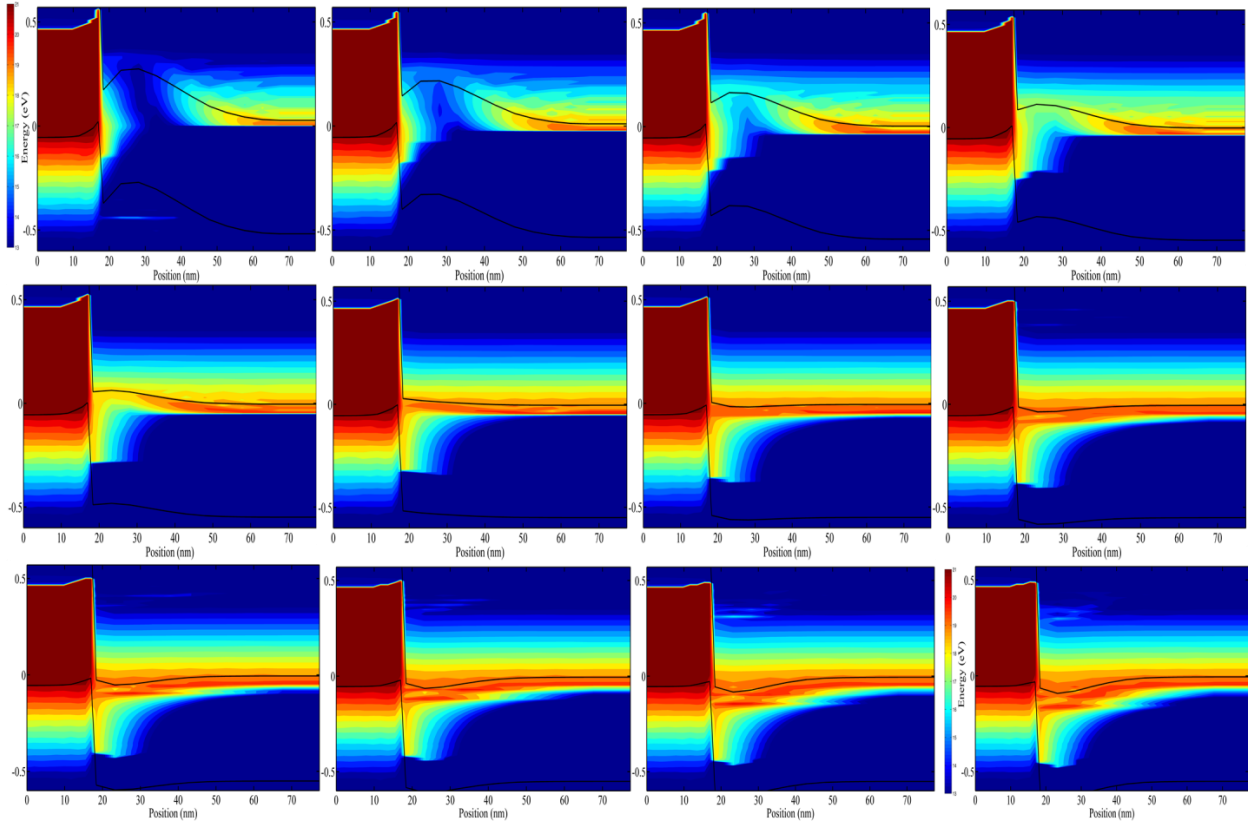

Fig.6. Energy-position resolved LDOS (x, E) and energy-position resolved electron density spectrum $G_n$ (x, E) distribution in the OFF-state $p^+$-$Al_{0.3}Ga_{0.7}Sb$ Pocket-Implanted GaSb/InAs TFETs

B. Current-Voltage characteristics

Figure 7 shows the $I_{DS}$-$V_{GS}$ transfer characteristics of $p^+$-$Al_{0.3}Ga_{0.7}Sb$ Pocket-Implanted GaSb/InAs TFETs on different value of $V_{DS}$ at 0.3V and 0.03V. On applying Gate voltage $V_{GS}$ of 0.6 V in the on state at drain voltage $V_{DS}$ of 0.3 V, an $I_{ON}$ current of 1A/μm and an $I_{ON}/I_{OFF}$ ratio of more than $10^8$ with subthreshold swing of about 32.6mV/decade are obtained. For $V_{DS}$ = 0.3 V, we obtain lowest value of drain current at gate voltage, i.e., $V_{GS}$ = 0 V which indicates that the gate has good control on the tunnel junction. In the ON-state for gate voltage $V_{GS}$ higher than 0.7 V drain current almost saturates and TFET has high output resistance.

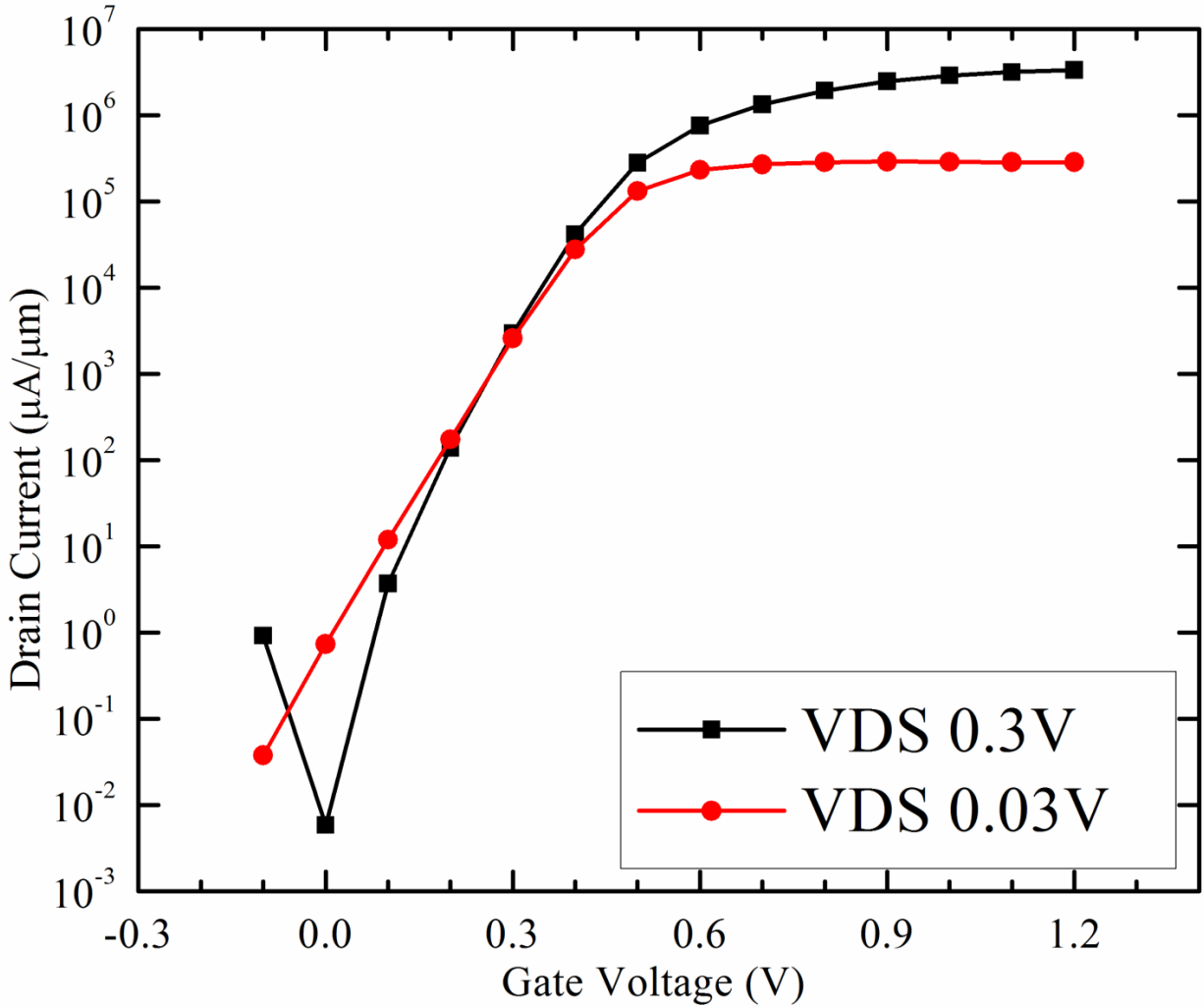

Fig.7. $I_{DS}$–$V_{GS}$ characteristics of the $p^+$-$Al_{0.3}Ga_{0.7}Sb$ Pocket-Implanted GaSb/InAs TFETs at $V_{DS}$= 0.03V and $V_{DS}$= 0.3V

Figure 8 shows the $I_{DS}$–$V_{DS}$ output characteristics of $p^+$-$Al_{0.3}Ga_{0.7}Sb$ Pocket-Implanted GaSb/InAs TFETs with Gate voltage variation $V_{GS}$ from 0.1 V to 1.2 V. For $V_{GS}$ of 0.1V and below, the off current is mainly due to thermionic emission leakage and the current due to tunnelling process is negligible. In the ON state at drain voltage $V_{DS}$ of 0.3 V the entire area of tunnelling junction is turned on and current density is uniform across the junction. Figure 9 shows the ON-State $I_{DS}$–$V_{GS}$ transfer characteristics of $p^+$-$Al_{0.3}Ga_{0.7}Sb$ Pocket-Implanted GaSb/InAs TFETs with variation in temperature. Temperature dependence is found to be extremely minute from positive gate voltages of 0 V to 1.2 V in the temperature range of 200K to 400K and hence this clearly indicates direct band-to-band tunnelling. For positive gate voltages of 0.7 V and above, the temperature dependence is extremely minute and current tends to saturate.

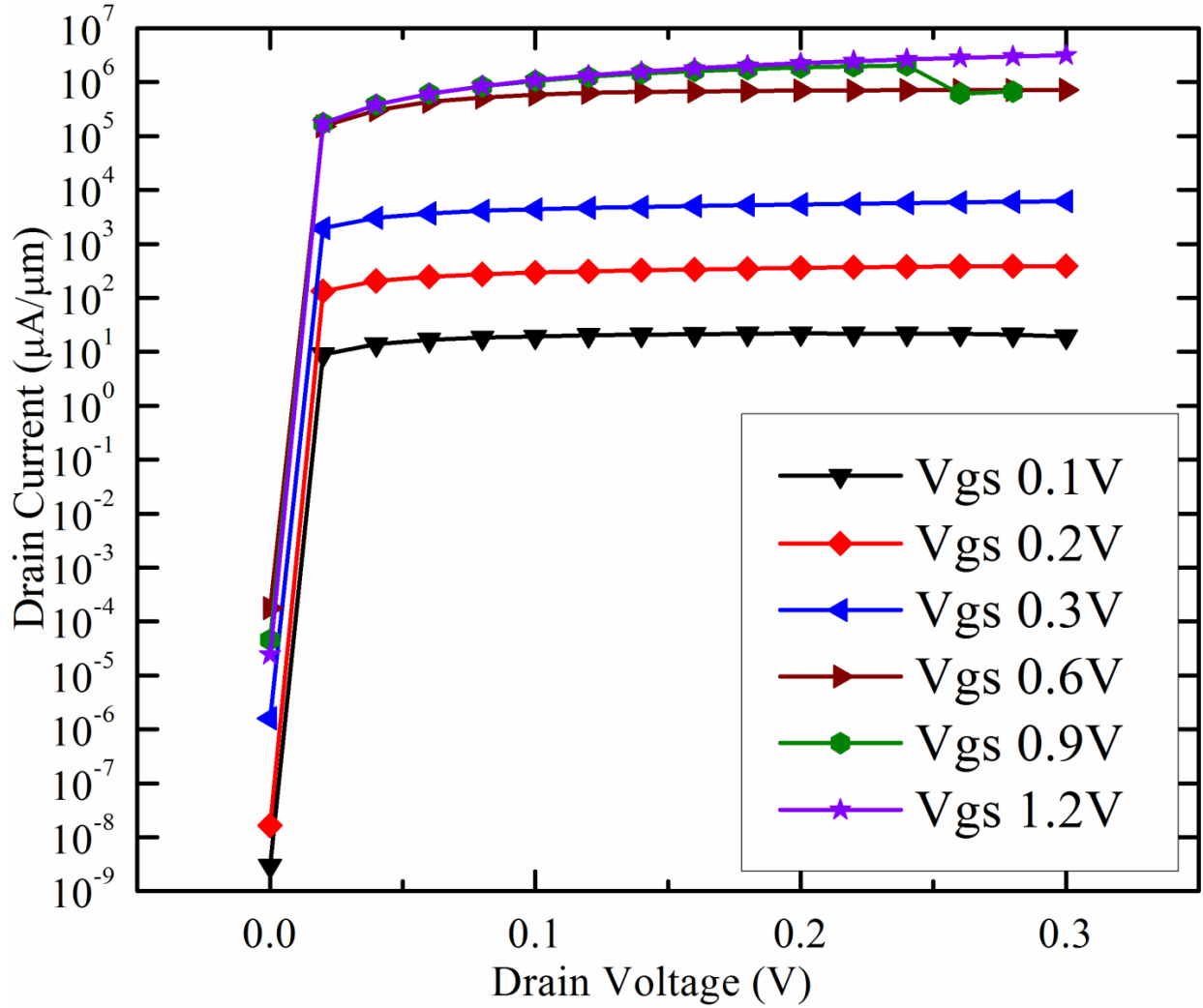

Fig.8. $I_{DS}$–$V_{DS}$ output characteristics of $p^+$-$Al_{0.3}Ga_{0.7}Sb$ Pocket-Implanted GaSb/InAs TFETs with Gate voltage variation $V_{GS}$

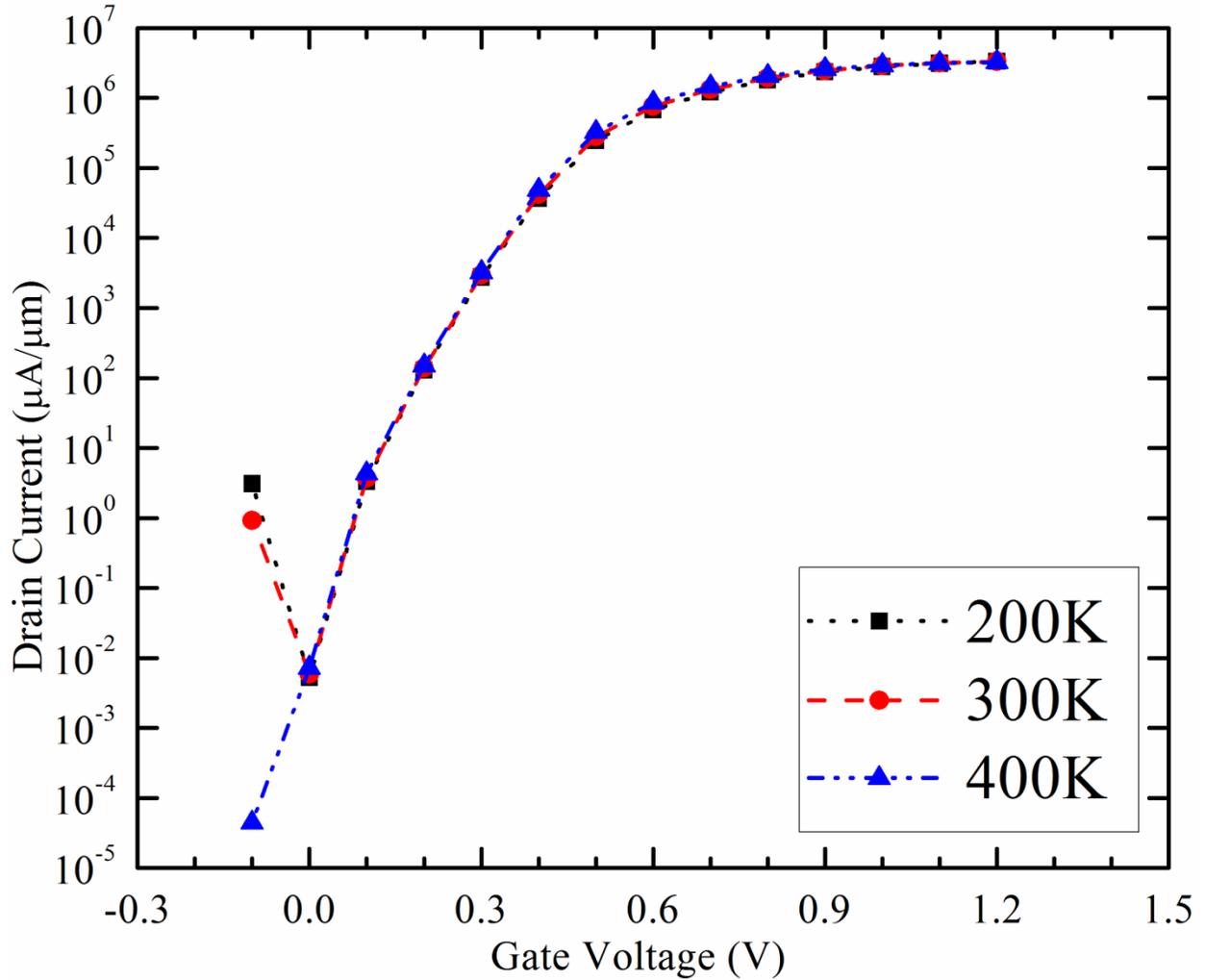

Fig.9. $I_{DS}$–$V_{GS}$ characteristics of the p$^+$-Al$_{0.3}$Ga$_{0.7}$Sb Pocket-Implanted GaSb/InAs TFETs at $V_{DS}$= 0.3V with variation in Temperature

## C. Effect of Geometry variation in p$^+$-Al$_{0.3}$Ga$_{0.7}$Sb Pocket-Implanted GaSb/InAs TFETs

We have investigated five different geometric variations in this section. Figure 10 shows ON-state $I_{DS}$–$V_{GS}$ transfer characteristics of the p$^+$-Al$_{0.3}$Ga$_{0.7}$Sb Pocket-Implanted GaSb/InAs TFETs with variation in drain length ($L_D$) extensions. The ambipolar current is estimated in the real devices sense since the simulation includes the quantization effect and effective increase in channel bandgap. Electrostatic effect remains unaffected by the channel quantization and increase in channel bandgap. We observed that increasing the drain length ($L_D$) extension increases the ambipolar conduction current and hence turn-on characteristics and $I_{ON}$ current largely increases.

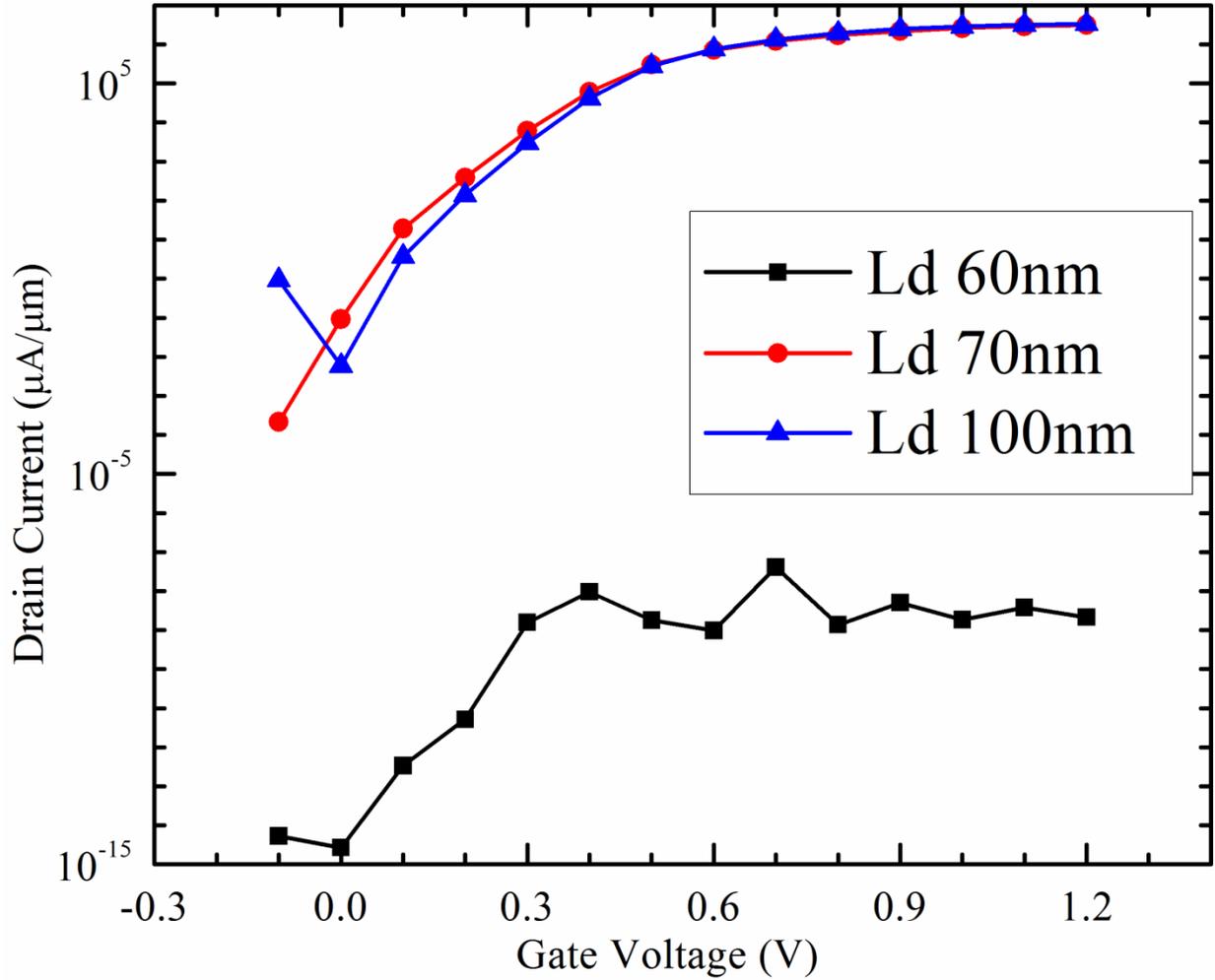

Fig.10. $I_{DS}$–$V_{GS}$ characteristics of the $p^+$-$Al_{0.3}Ga_{0.7}Sb$ Pocket-Implanted GaSb/InAs TFETs at $V_{DS}$= 0.3V with variation in drain length ($L_D$)

Figure 11 shows the $I_{DS}$–$V_{GS}$ characteristics of the $p^+$-$Al_{0.3}Ga_{0.7}Sb$ Pocket-Implanted GaSb/InAs TFETs at $V_{DS}$= 0.3 V with variation in undercut lengths ($L_{UC}$). At zero undercut lengths, $I_{ON}/I_{OFF}$ ratio is reduced by five orders of magnitude and the subthreshold swing increases to 130.4 mV/decade. On increasing undercut length to 5 nm subthreshold swing reduces to 65.1 mV/decade and further at undercut length of 10 nm, subthreshold swing reduces to the 32.6 mV/decade and $I_{ON}/I_{OFF}$ ratio increases. $I_{ON}$ current reduces on further increases in undercut lengths as $I_{ON}$ current is proportional to the area of tunnel junction and hence on increasing the undercut lengths ($L_{UC}$) width-normalized $I_{ON}$ current density decreases. Here the optimization criterion is overlapping the gate on the tunnelling region. The scalability of $p^+$-$Al_{0.3}Ga_{0.7}Sb$ Pocket-Implanted GaSb/InAs TFETs is limited by undercut lengths ($L_{UC}$) which is necessary to achieve a steep slope.

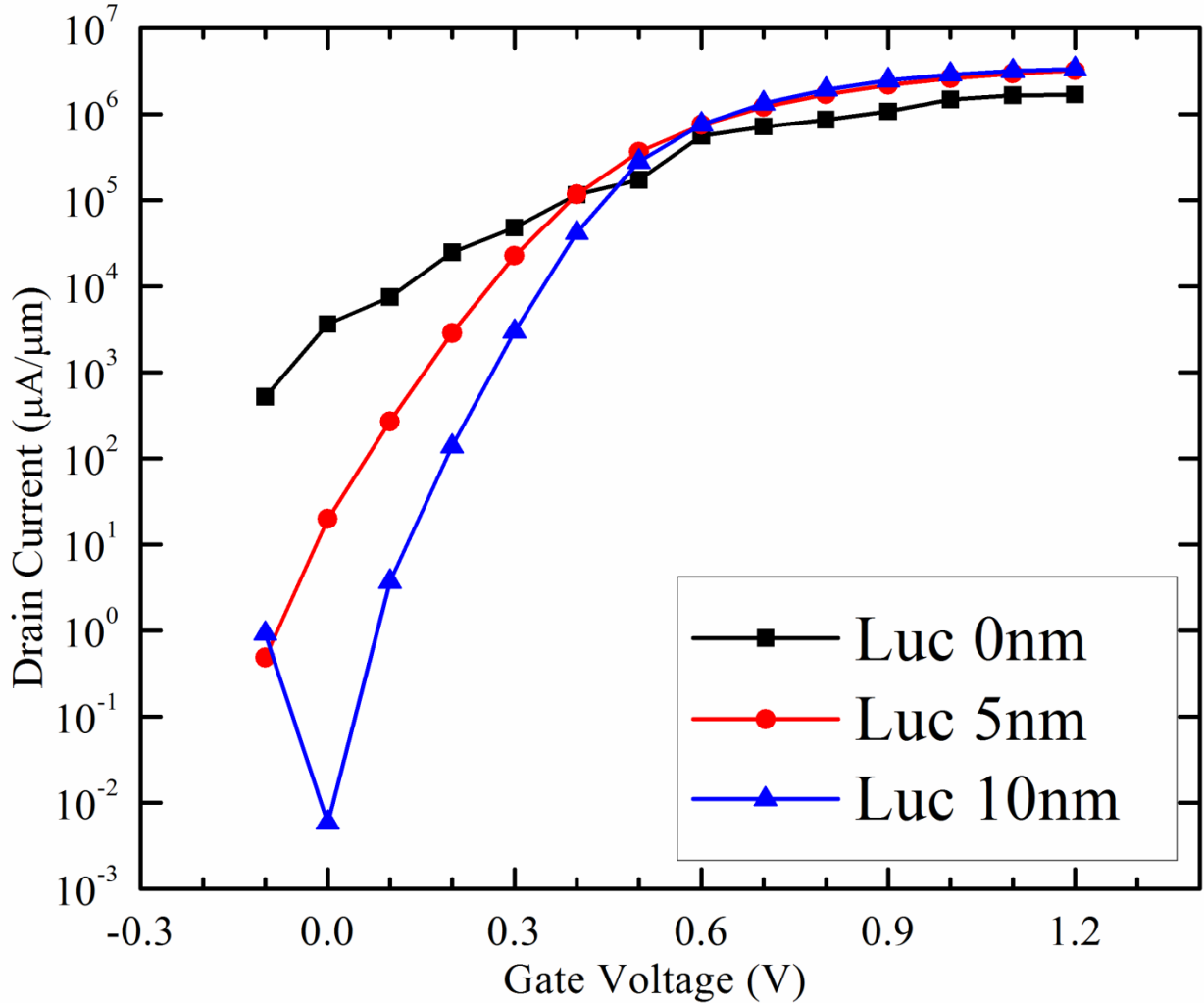

Fig.11. $I_{DS}$–$V_{GS}$ characteristics of the $p^+$-$Al_{0.3}Ga_{0.7}Sb$ Pocket-Implanted GaSb/InAs TFETs at $V_{DS}$=0.3V with variation in undercut lengths ($L_{UC}$)

In the figure 12 we showed the variation of equivalent oxide thickness ($T_{OX}$) on $p^+$-$Al_{0.3}Ga_{0.7}Sb$ Pocket-Implanted GaSb/InAs TFETs. A thinner $T_{OX}$ gives strong coupling between InAs channel and gate field and hence gives rise to steeper subthreshold swing. With variation in $T_{OX}$ from 1.2 nm to 4 nm subthreshold swing increases from 32.6 mV/decade to 58 mV/decade. In the thinner $T_{OX}$ due to strong coupling ambipolar current also increases and hence, $I_{ON}$ current slightly increases. For the $p^+$-$Al_{0.3}Ga_{0.7}Sb$ Pocket-Implanted GaSb/InAs TFETs of given undercut lengths ($L_{UC}$) subthreshold slope depends upon the $T_{OX}$. In the figure 13 we showed the variation of different Gate oxide material on $p^+$-$Al_{0.3}Ga_{0.7}Sb$ Pocket-Implanted GaSb/InAs TFETs. For $Al_2O_3$ High-K Gate material subthreshold swing reduces to 31mV/decade and for $HfO_2$ High-K Gate material subthreshold swing is 24.1mV/decade.

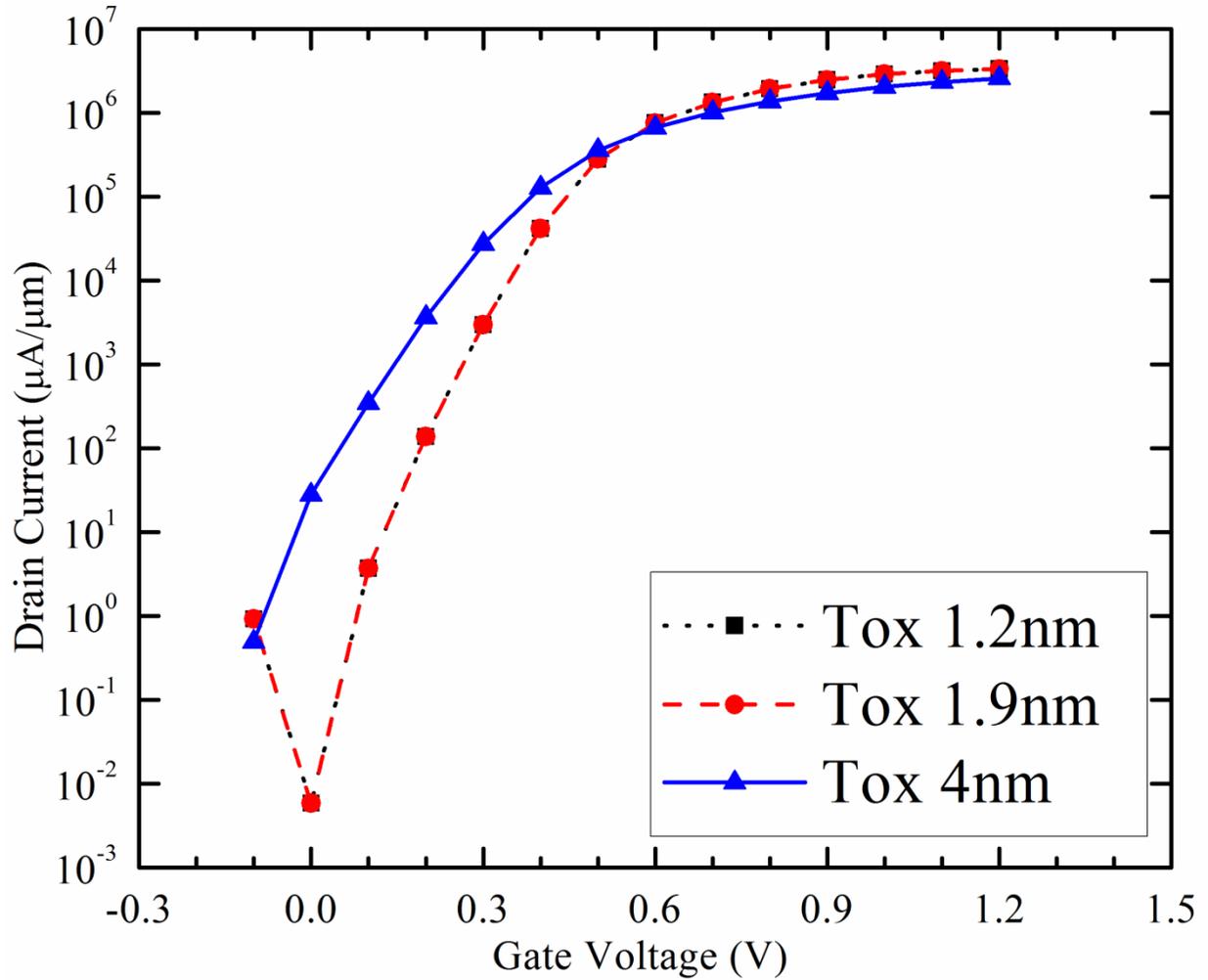

Fig.12. $I_{DS}$–$V_{GS}$ characteristics of the $p^+$-$Al_{0.3}Ga_{0.7}Sb$ Pocket-Implanted GaSb/InAs TFETs at $V_{DS}$= 0.3V with variation in Tox

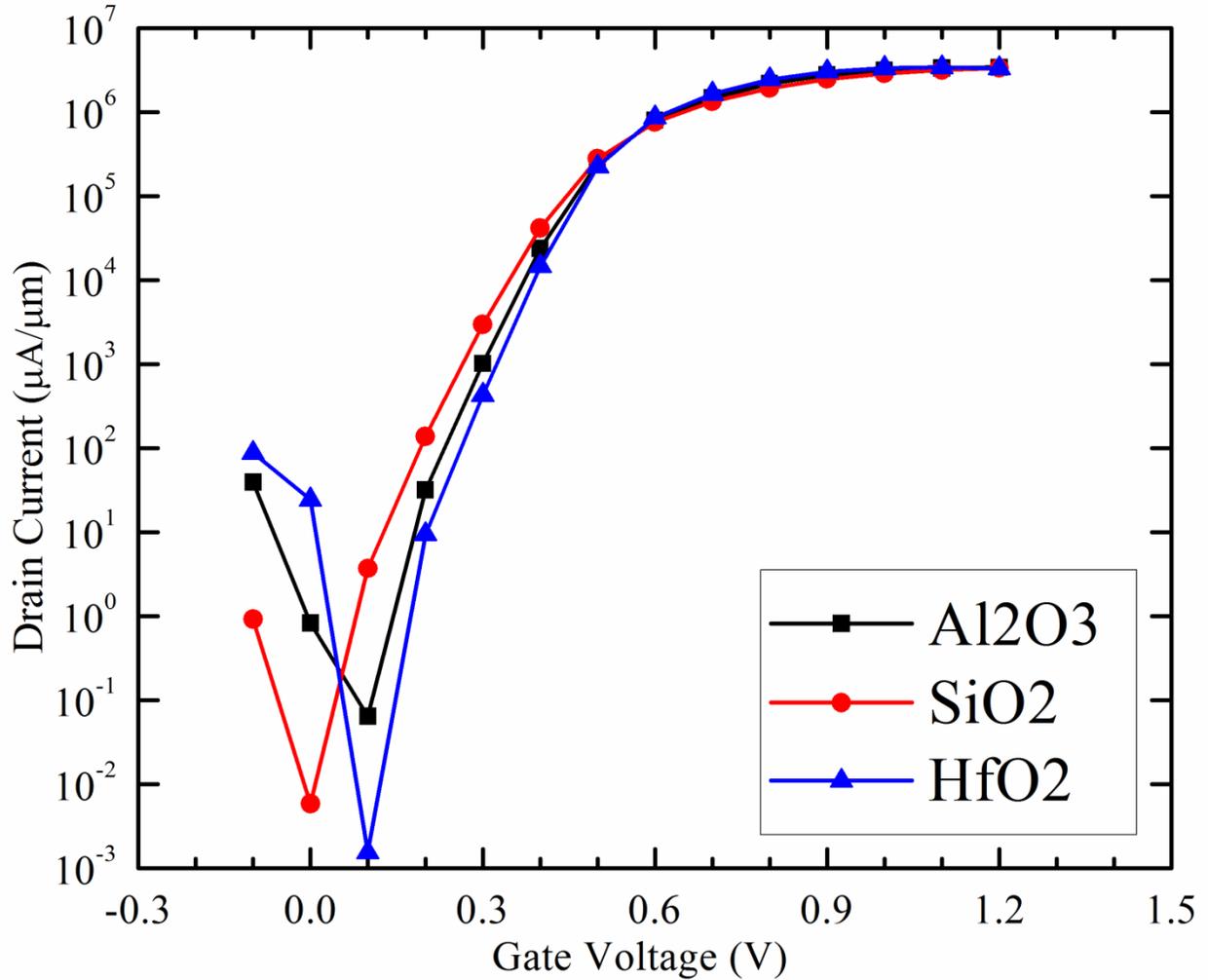

Fig.13. $I_{DS}$–$V_{GS}$ characteristics of the $p^+$-$Al_{0.3}Ga_{0.7}Sb$ Pocket-Implanted GaSb/InAs TFETs at $V_{DS}$= 0.3V with variation in Gate Oxide Material

In the figure 14, we showed the effect of variation in gate-length ($L_G$) on $p^+$-$Al_{0.3}Ga_{0.7}Sb$ Pocket-Implanted GaSb/InAs TFETs. We are varying the gate lengths ($L_G$) from 10 nm to 30 nm while holding rest of all the geometric parameters constant and comparing the $I_{DS}$–$V_{GS}$ transfer characteristics. Gate length ($L_G$) had a weak influence on $I_{ON}$ current. But, as gate length ($L_G$) increases overlap above tunnelling junction area also increases and hence gate has better control to shut off $I_{OFF}$ current effectively. Hence, $I_{OFF}$ current decreases. $I_{OFF}$ current decreases with the increase of gate length ($L_G$) from 10 nm to 25 nm. Gate length ($L_G$) also strongly influences the minimum value of $I_{OFF}$ current and hence subthreshold swing. As gate length increases minimum value of $I_{OFF}$ current also decreases with increase in subthreshold swing as shown in Figure 14.

For gate-length of 30 nm onset of cut off voltage is at $V_{GS}$ 0.1V and for gate-length of 25 nm onset of cut off voltage is at $V_{GS}$ 0V and on further decrease in gate-length to 20 nm onset of cut off voltage is at $V_{GS}$ -0.1 V. These simulation results suggest that the device performance of $p^+$-$Al_{0.3}Ga_{0.7}Sb$ Pocket-Implanted GaSb/InAs TFETs is determined by gate-drain periphery and gate has good control on the center of the tunnel junction to turn off the devices.

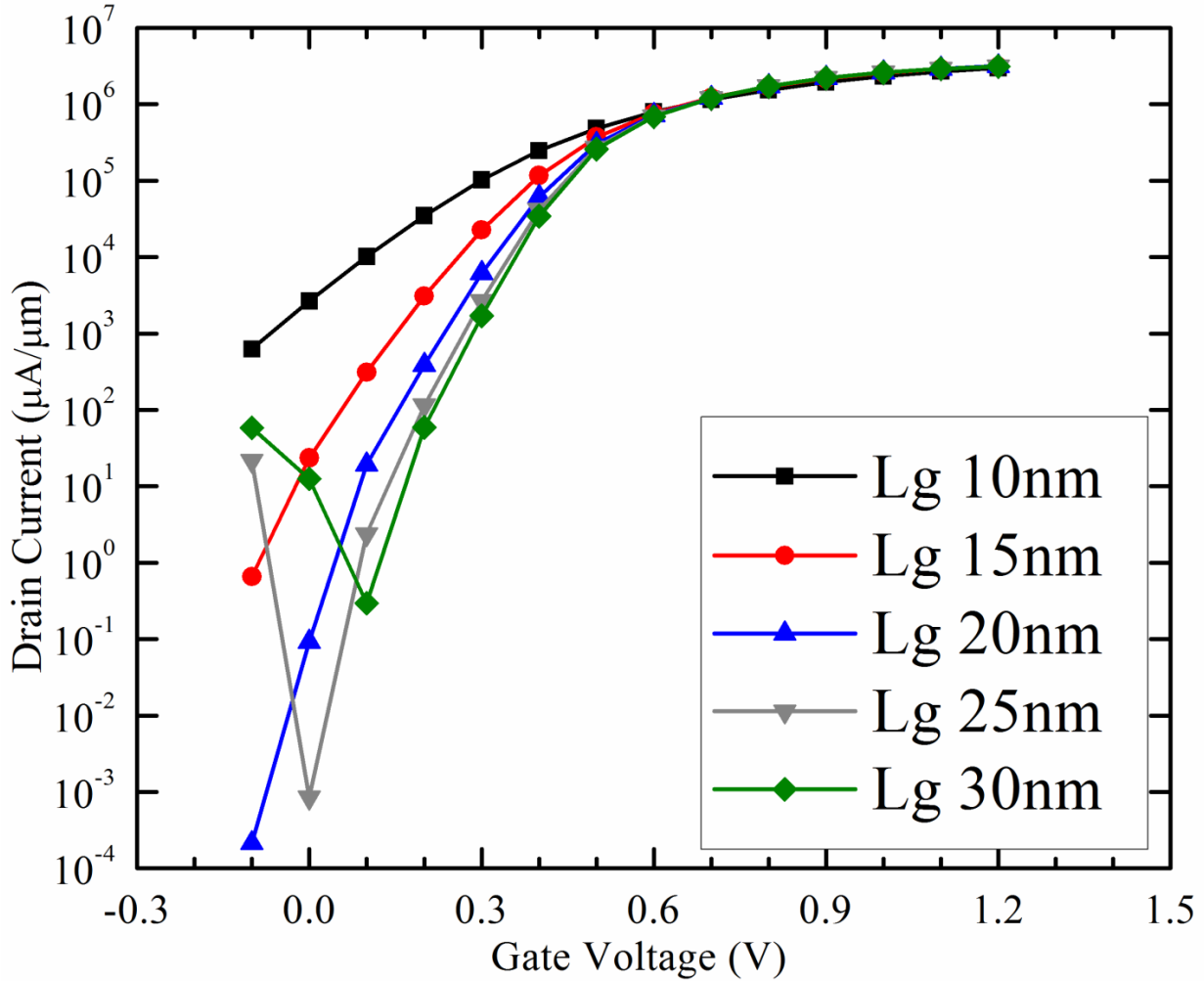

Fig.14. $I_{DS}$–$V_{GS}$ characteristics of the $p^+$-$Al_{0.3}Ga_{0.7}Sb$ Pocket-Implanted GaSb/InAs TFETs at $V_{DS}$= 0.3V with variation in Gate Length ($L_G$)

In the figure 15, we showed the effect of variation of Drain Thickness ($T_{InAs}$) in $p^+$-$Al_{0.3}Ga_{0.7}Sb$ Pocket-Implanted GaSb/InAs TFETs. For the $p^+$-$Al_{0.3}Ga_{0.7}Sb$ Pocket-Implanted GaSb/InAs TFETs of a given undercut length ($L_{UC}$) $I_{OFF}$ current and subthreshold slope depend upon the InAs channel thickness ($T_{InAs}$). For larger Drain Thickness ($T_{InAs}$) Gate will lose the control on tunnel junction to shut off $I_{OFF}$ current effectively. We have observed that at gate voltage of 0.6 V

$I_{ON}$ current increases by more than an order of magnitude when Drain Thickness ($T_{InAs}$) increases from 3 nm to 4 nm but with the overhead of higher value of subthreshold slope. For Drain Thickness ($T_{InAs}$) of 7 nm, we have observed that at gate voltage of 0.5 V $I_{ON}$ current of as high as 1 A/μm is achieved with $I_{OFF}$ of 0.5mA/μm and subthreshold slope of 65.2 mV/decade. We also observe that high $I_{ON}$ current can be achieved by higher InAs concentration but at the cost of high value of subthreshold slope.

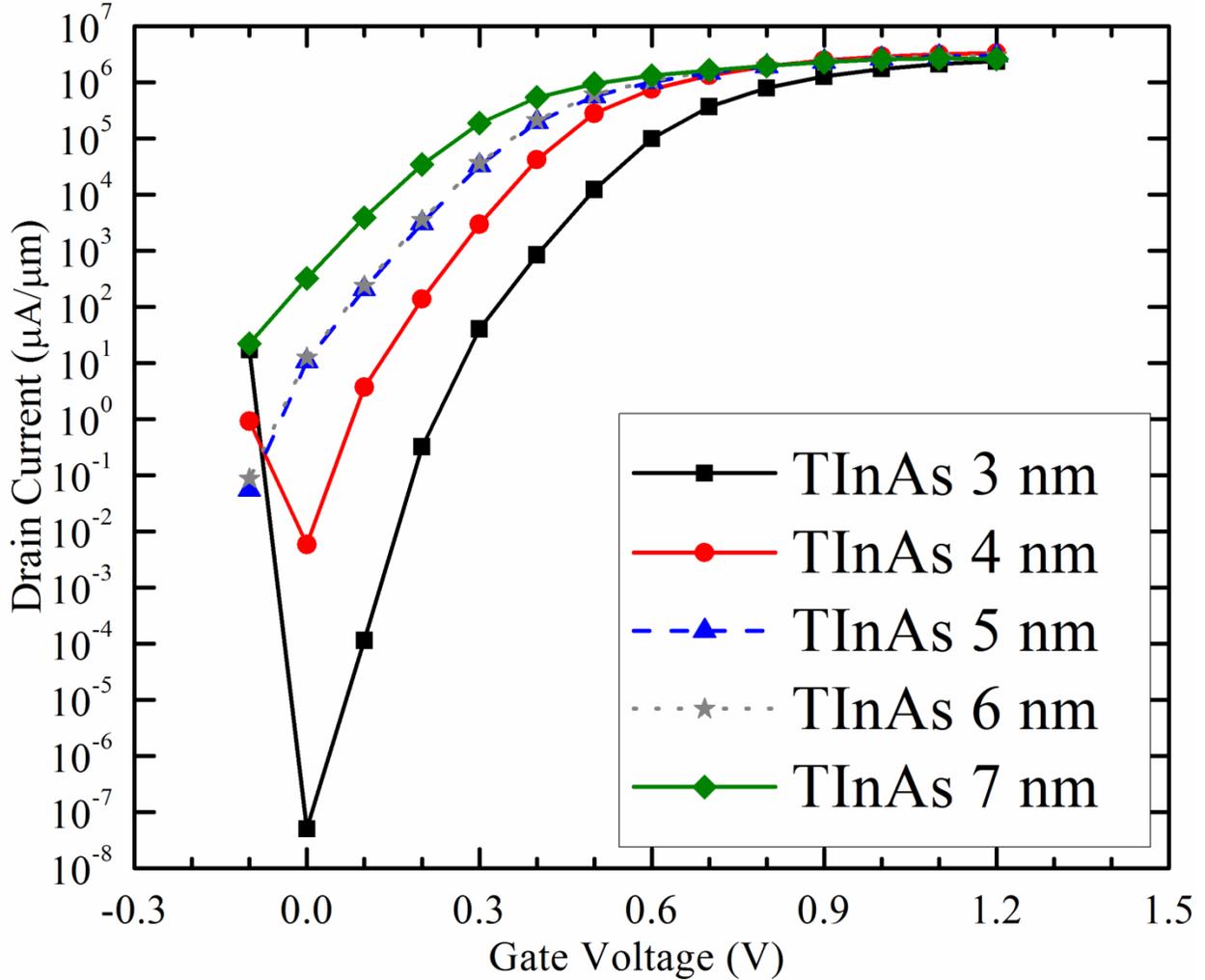

Fig.15. $I_{DS}$–$V_{GS}$ characteristics of the $p^+$-$Al_{0.3}Ga_{0.7}Sb$ Pocket-Implanted GaSb/InAs TFETs at $V_{DS}$= 0.3 V with variation in Drain Thickness ($T_{InAs}$)

In the figure 16, we showed $I_{DS}$–$V_{GS}$ characteristics of the Pocket-Implanted GaSb/InAs TFETs at $V_{DS}$= 0.3 V without and with $p^+$-GaSb Pocket Implant and $p^+$-$Al_{0.3}Ga_{0.7}Sb$ Pocket Implant. The effect of $p^+$-$Al_{0.3}Ga_{0.7}Sb$ Pocket-Implant is clearly visible with increase in $I_{ON}$ current by 3 order of magnitude at gate voltage of 0.6 V but the SS decreases from 11.85 mV/decade to 32.6 mV/decade.

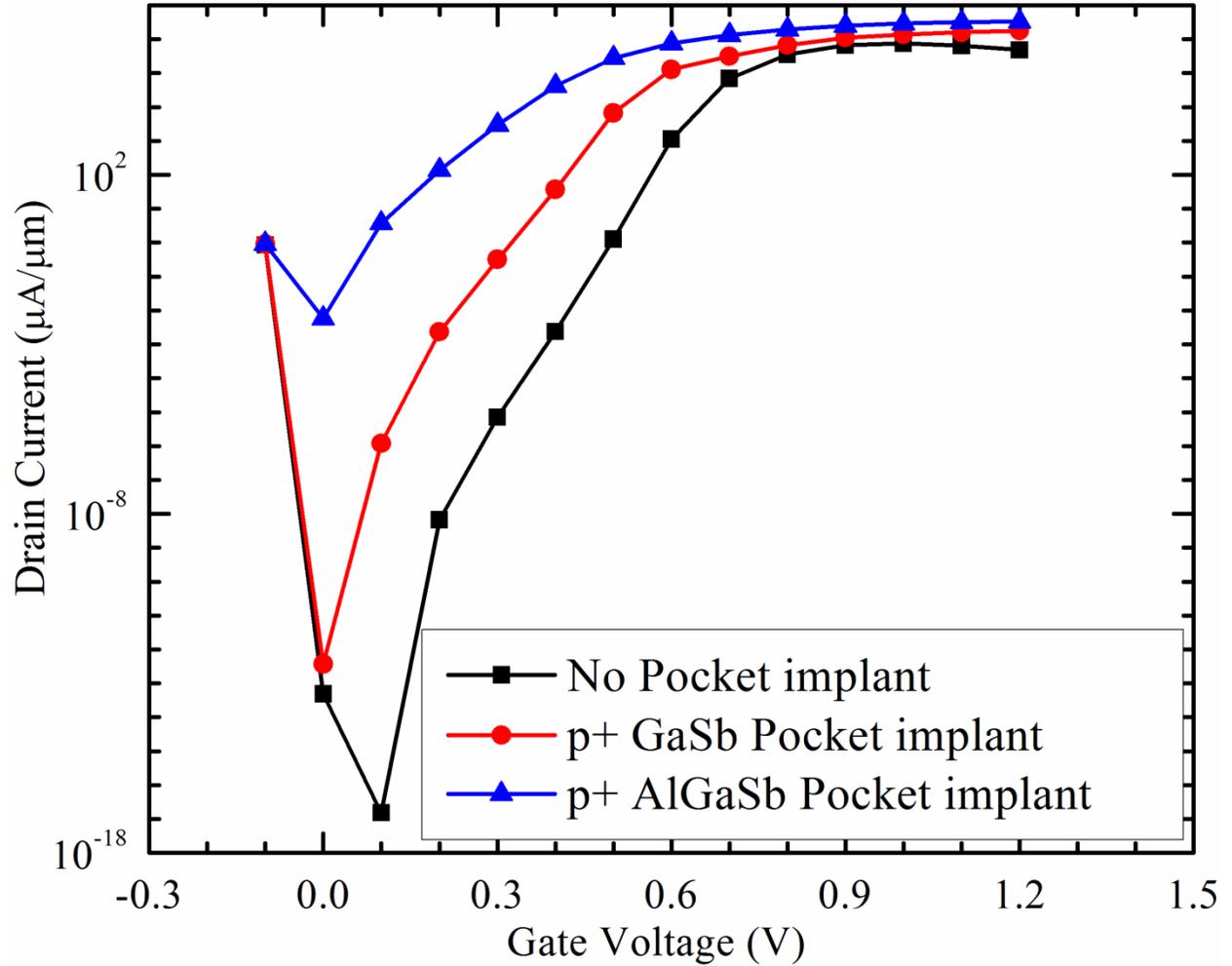

Fig.16. $I_{DS}$–$V_{GS}$ characteristics of the GaSb/InAs TFETs at $V_{DS}$= 0.3 V without and with $p^+$-GaSb Pocket-Implant and with $p^+$-$Al_{0.3}Ga_{0.7}Sb$ Pocket Implant

CONCLUSION

In this work, we used a method based on combined semiclassical electrostatic potential and NEGF calculation on a 3-D full-band atomistic $sp^3d^5s^*$ spin-orbital coupled tight-binding based quantum mechanical simulator and applied to $p^+$-$Al_{0.3}Ga_{0.7}Sb$ Pocket-Implanted L-shaped GaSb/InAs staggered-bandgap (SG) heterojunction vertical n-channel tunnel field-effect transistors (TFETs) of 4 nm thin channel structures with gate length of 20 nm. Advantage of this geometry is that the gate electric field is in line with the tunnelling current. Simulation results show that $p^+$-$Al_{0.3}Ga_{0.7}Sb$ Pocket-Implanted GaSb/InAs n-TFETs are capable of low-voltage operation. On applying Gate voltage $V_{GS}$ of 0.6 V, at the on state of drain voltage $V_{DS}$ of 0.3 V, an $I_{ON}$ current of 1A/μm, an $I_{ON}/I_{OFF}$ ratio of more than $10^8$ with subthreshold swing of about 32.6 mV/decade are obtained. We also study the effect of variation of different geometry on I-V characteristics of $p^+$-$Al_{0.3}Ga_{0.7}Sb$ Pocket-Implanted GaSb/InAs n-TFETs. These simulation results suggest that, the device performance of $p^+$-$Al_{0.3}Ga_{0.7}Sb$ Pocket-Implanted GaSb/InAs TFETs is determined by drain-gate periphery and gate has good control on the center of tunnelling junction to turn off the devices. In the TFETs only at the junctions high electric fields exist and current is largely determined by screened tunnelling length. Hence, gate length scaling rules of MOSFETs are not eligible for TFETs and after the leakage becomes predominant, length of intrinsic region has very negligible effect on device performances. In conclusion, simulation results shown in this work demonstrate that Pocket-Implanted L-shaped GaSb/InAs staggered-bandgap (SG) heterojunction vertical n-TFET is the versatile building block for future low power electronics and it is reasonable to experimentally investigate these TFET geometries.